\documentclass[conference]{IEEEtran}

\IEEEoverridecommandlockouts
\usepackage{cite}
\usepackage{amsmath,amssymb,amsfonts}
\usepackage{graphicx}
\usepackage{textcomp}
\usepackage{xcolor}
\def\BibTeX{{\rm B\kern-.05em{\sc i\kern-.025em b}\kern-.08em
    T\kern-.1667em\lower.7ex\hbox{E}\kern-.125emX}}
\usepackage{xspace}
\usepackage{xcolor}
\usepackage{enumitem}
\usepackage{algorithm}
\usepackage{algpseudocode}
\usepackage{graphicx}
\usepackage{subcaption}
\usepackage{caption}
\usepackage{booktabs}

\newcommand{\SysName}{{\scshape CheckMate}\xspace}
\newcommand{\ChainReactor}{{\scshape ChainReactor}\xspace}
\newcommand{\PentestGPT}{{\scshape PentestGPT}\xspace}
\newcommand{\PentestAgent}{{\scshape PentestAgent}\xspace}
\newcommand{\CAI}{{CAI}\xspace}
\newcommand{\AutoAttacker}{{\scshape AutoAttacker}\xspace}
\newcommand{\AutoPentester}{{\scshape AutoPentester}\xspace}
\newcommand{\PenHeal}{{\scshape PenHeal}\xspace}
\newcommand{\VulnBot}{{\scshape VulnBot}\xspace}

\begin{document}

\title{Automated Penetration Testing with LLM Agents and Classical Planning}



\author{
\IEEEauthorblockN{
Lingzhi Wang\IEEEauthorrefmark{1},
Xinyi Shi\IEEEauthorrefmark{1},
Ziyu Li\IEEEauthorrefmark{1},
Yi Jiang\IEEEauthorrefmark{2},
Shiyu Tan\IEEEauthorrefmark{2},
Yuhao Jiang\IEEEauthorrefmark{1},
Junjie Cheng\IEEEauthorrefmark{2},
Wenyuan Chen\IEEEauthorrefmark{2},\\
Xiangmin Shen\IEEEauthorrefmark{3},
Zhenyuan LI\IEEEauthorrefmark{2},
Yan Chen\IEEEauthorrefmark{1}}
\IEEEauthorblockA{
\IEEEauthorrefmark{1}\textit{Northwestern University}, 
\IEEEauthorrefmark{2}\textit{Zhejiang University},
\IEEEauthorrefmark{3}\textit{Hofstra University}
}
}

\maketitle

\begin{abstract}
While penetration testing plays a vital role in cybersecurity, achieving fully automated, hands-off-the-keyboard execution remains a significant research challenge.
In this paper, we introduce the ``Planner-Executor-Perceptor (PEP)'' design paradigm and use it to systematically review existing work and identify the key challenges in this area.
We also evaluate existing penetration testing systems, with a particular focus on the use of Large Language Model (LLM) agents for this task.
The results show that the out-of-the-box Claude Code and Sonnet 4.5 exhibit superior penetration capabilities observed to date, substantially outperforming all prior systems.
However, a detailed analysis of their testing processes reveals specific strengths and limitations; notably, LLM agents struggle with maintaining coherent long-horizon plans, performing complex reasoning, and effectively utilizing specialized tools.
These limitations significantly constrain its overall capability, efficiency, and stability.
To address these limitations, we propose \SysName, a framework that integrates enhanced classical planning with LLM agents, providing an external, structured ``brain" that mitigates the inherent weaknesses of LLM agents.
Our evaluation shows that \SysName outperforms the state-of-the-art system (Claude Code) in penetration capability, improving benchmark success rates by over 20\%.
In addition, it delivers substantially greater stability, cutting both time and monetary costs by more than 50\%.
\end{abstract}


\section{Introduction}
Penetration testing, hereafter used interchangeably with ``pentesting,'' has become a vital component of cybersecurity.
It enables organizations to proactively identify and mitigate vulnerabilities before exploited by adversaries.
The U.S. Cybersecurity and Infrastructure Security Agency (CISA) highlights pentesting as a core service for protecting critical infrastructure~\cite{CISA2023}.
MarketsandMarkets~\cite{MarketsandMarkets2024} forecasts that the global pentesting market was valued at USD 1.7 billion in 2024, and is projected to grow to USD 3.9 billion by 2029~\cite{MnM_PTaaS_2024}.
The PenTesting-as-a-Service (PTaaS) segment alone is expected to expand from USD 118 million in 2024 to USD 301 million by 2029.
This significant and rising demand has prompted the adoption of artificial intelligence (AI) to meet the need for more efficient and scalable penetration testing solutions.
A recent research~\cite{mayoralvilches2025caiopenbugbountyready} suggests that by 2028, AI-powered security testing tools will outnumber human pentesters in mainstream security operations, signaling a transformative shift in the field.

The high reliability of skilled human pentesters has made fully automated penetration testing a long-standing challenge in cybersecurity research.
However, existing approaches still require nontrivial human intervention and have yet to achieve truly ``hands-off-the-keyboard'' automation~\cite{wu2024autopt}.
As a result, users are often forced to act as supervisors, executors, prompters, and decision-makers, limiting the practical utility of these systems.
Moreover, human intervention hinders rigorous and unbiased evaluation, as it is difficult to standardize the extent of human involvement during evaluation, leading to inconsistent assessments across studies.
In real-world use, different human knowledge, experience, and skill levels make it harder for the system to achieve its expected effectiveness.

Multiple challenges have prevented truly ``hands-off-the-keyboard" automation in pentesting.
Earlier work based on formal models~\cite{sarraute2012pomdps,hoffmann2015simulated} is unable to interpret heterogeneous, unstructured information or generate specific commands.
Therefore, they either operate on overly simplified scenarios or produce abstract attack plans without concrete commands or instructions.
Recent advances in LLMs enable heterogeneous information processing and detailed instruction generation, yet human involvement remains necessary in LLM-based pentesting.
First, LLM hallucinations often lead to incomplete, incorrect, or even fabricated commands, requiring human supervision to identify and correct them before execution~\cite{wu2024autopt}.
Second, limited context memory and logical reasoning capablilties~\cite{mirzadeh2024gsm,lin2025zebralogic,yamin2024failure,chi2024unveiling} make LLMs ineffective at performing long-term, multi-step pentesting tasks~\cite{wu2024autopt}.
They often become stuck in repetitive and unproductive iterations and require human input and guidance.

Recently, LLMs have been widely applied to complex tasks such as software development~\cite{liu2024large,jin2024llms,wang2025agents}, code auditing~\cite{ullah2024llms,guo2025repoaudit}, and data engineering~\cite{rahman2025llm}, prompting a focus on \textit{LLM Agents}.
These systems are characterized by their ability to automatically plan for complex tasks, execute code and commands, and iteratively refine them based on execution results, significantly reducing the need for human guidance and intervention.
In this paper, we first try to answer the question: \textit{Can LLM agents conduct pentesting independently?}
To answer this, we evaluated and compared their penetration capabilities against prior work.
Our results show that the out-of-the-box Claude Code~\cite{anthropic_claude_code} and Sonnet 4.5 can autonomously complete pentesting tasks and achieve state-of-the-art (SOTA) capability, substantially outperforming all prior work in this field.
To understand the strong performance of Claude Code, we further analyze its testing process and find that it demonstrates superior code refinement and subtask management capabilities.
However, we also identify several limitations in Claude Code and other LLM agents in pentesting, including difficulties in maintaining coherent long-horizon plans, performing complex reasoning, and leveraging specialized tools.
Such issues lead to lower success rates, efficiency, and unstable performance.

In this paper, we address these limitations by integrating classical planning techniques with LLM agents.
Our approach is motivated by several key insights.
First, classical planning represents plans as directed acyclic graphs (DAGs)~\cite{ghallab2004automated}, which provide an explicit and logically structured view of the overall task.
This structure enables the agent to maintain a coherent long-horizon plan and mitigates common failure modes, such as erratic attack-path selection, incomplete or repetitive execution, and forgetting previously gathered information.
Second, classical planning explicitly encodes causal relationships through preconditions and effects.
This allows us to generate long reasoning chains without relying on the LLM itself, thereby ensuring their correctness while maintaining efficient and stable reasoning.
Moreover, this explicit causal structure enables us to incorporate customized domain knowledge and reasoning patterns that go beyond the LLM’s internal knowledge.
Last but not least, the use of modular actions defined in classical planning allows us to easily integrate uncommon or highly specialized pentesting tools, reducing reliance on the LLM’s internal knowledge for command generation.
These predefined actions also let us specify critical execution details in advance, eliminating the need for repeated LLM iterations to refine simple commands on the fly.
This design further improves accuracy, efficiency, and overall system stability.

However, traditional classical planning is limited to fully observable and deterministic environments, making it unsuitable for pentesting scenarios.
To address this gap, we propose \textit{Classical Planning+}, the first LLM-augmented classical planning framework capable of dynamic updates.
Classical planning+ preserves the core principles of classical planning while leveraging LLMs to update action effects and state information on the fly, thereby removing the requirement to fully specify all domain knowledge beforehand.
As a result, it extends classical planning to partially observable and non-deterministic domains, significantly broadening its applicability to real-world pentesting tasks.

In this paper, we aim to enhance LLM-based pentesting with classical planning+.
We decompose a pentesting system into three core components: a planner, an executor, and a perceptor.
Using this design paradigm, we review the design choices in existing work and introduce our system, \SysName.
In our system, classical planning+ serves as the planner: it infers the set of feasible actions given the current attack state and selects the most appropriate next step.
LLM agents act as the executor that carries out these actions.
Their execution outputs are then converted by an LLM into predicates compatible with classical planning+, enabling the planner to update the attack state and determine subsequent steps.
Our experiments on the Vulhub~\cite{vulhub} dataset demonstrate that \SysName significantly outperforms existing systems, including Claude Code, in penetration capability, efficiency, and stability.

Our main contributions are summarized as follows.
\begin{enumerate}[leftmargin=1.5em]
    \item \textbf{Proposed a Unified Design Paradigm for Pentesting.} We propose a design paradigm for automated pentesting systems, which consists of three fundamental components: a Planner, an Executor, and a Perceptor (PEP).
    We use this paradigm to review and categorize existing work.
    This paradigm helps us understand the distinctions among existing works and provides guidance for the future development and improvement of such systems.
    \item{\textbf{Performed the Largest Evaluation on Existing Systems.}} We systematically evaluate existing pentesting systems on the Vulhub dataset.
    We find that out-of-the-box Claude Code, powered by Sonnet 4.5, achieves the strongest performance with the least human intervention, showing a substantial improvement over all prior work in this area.
    We further analyzed the penetration workflow of Claude Code and identified three major limitations.
    \item \textbf{Proposed Classical Planning+ and Developed \SysName.} To address these limitations, we propose classical planning+, the first dynamically updating classical planning scheme powered by LLMs, which extends conventional classical planning to partially observable and non-deterministic tasks.
    Based on it, we develop \SysName, which integrates classical planning and LLM agents for pentesting.
    Our extensive evaluation demonstrates that \SysName substantially outperforms prior systems, improving success rates on Vulhub benchmark by over 20\% and cutting both time and monetary costs by more than 50\%.
\end{enumerate}

\section{The PEP Paradigm and Related Work}
\subsection{The PEP Designing Paradigm for Pentesting Systems}\label{sec:pep-model-and-related-work}
With the growing volume of work in automated pentesting, we consolidate a design paradigm that decomposes an automated pentesting system into three cooperating components: a Planner, an Executor, and a Perceptor (PEP).
This decomposition provides a clear way to dissect systems and a framework for future research, allowing each component to be independently analyzed, improved, and benchmarked.
Using the PEP paradigm, we taxonomized prior work and, for each component, reviewed representative solutions, remaining challenges, and open problems.
The prior work taxonomy is summarized in Table~\ref{tab:pentesting-systems}.

\subsubsection{Planner}
The planner aims to answer: (1) what actions are feasible now; (2) among those feasible actions and their following attack paths, which one has the highest value and should be given priority for execution.
Earlier work used formal methods as the planner.
For example, some work~\cite{sarraute2012pomdps,schwartz2020pomdp+,sarraute2013penetration,schwartz2019autonomous,ghanem2023hierarchical} presents pentesting as a partially observable Markov decision process (POMDP) to model the uncertainty and incomplete information, where the planner must choose between information-gathering (e.g., scans) and offensive (e.g., exploit) actions, update a Bayesian-style belief state from observations, and enumerate feasible actions under the current belief state.
Follow-up research has extended POMDP to account for factors such as defender responses~\cite{schwartz2020pomdp+} or observation noise~\cite{schwartz2019autonomous,ghanem2023hierarchical,zhou2021autonomous}.
However, these systems suffer computational blow-up as the problem scales~\cite{sarraute2012pomdps}.
Moreover, their estimated probability models (e.g., scan success rates or exploitability probabilities) are difficult to obtain in the real world.
Another line of work~\cite{hoffmann2015simulated,depasquale_chainreactor,obes2013attack,chen2024survey,wang2021automatic} formulates planning as a path-search problem.
\ChainReactor~\cite{depasquale_chainreactor}, as a representative work, casts privilege-escalation attacks as a classical planning task.
However, those approaches treat pentesting as a static and deterministic problem, and thus do not address how to plan under those real-world uncertainties and incomplete information.

LLMs can be naturally used as a planner by supplying pentesting information as context and requesting the next action.
However, it suffers from some limitations, such as hallucinations, short-term memory, and limited context windows.
Therefore, existing work has designed various mechanisms to help LLMs plan.
For example, some systems maintain structured textual representations of the plan, including the Penetration Tree (PTT) and its variants in \PentestGPT~\cite{deng2024pentestgpt}, \AutoPentester~\cite{AutoPentestGPT}, and \PenHeal~\cite{huang2023penheal}, as well as the Situation Summaries in \AutoAttacker~\cite{xu2024autoattacker}.
VulnBot~\cite{kong2025vulnbot} leverages LLMs to convert the planning problem to a Penetration Task Graph (PTG).
\PentestAgent~\cite{PentestAgent} uses LLMs together with CVE/service keywords to drive a two-stage planning.
It leverages LLMs to interpret the reconnaissance results and extract CVEs or service names, which are used to search for exploits online.
Multi-agent systems like \CAI~\cite{mayoralvilches2025caiopenbugbountyready} rely on LLMs to coordinate available agents or tools.
In these frameworks, the LLM has function descriptions and pentesting information as context and determines which agent or tool to invoke at each step.
Some systems also maintain a to-do list to guide the agent’s progress and prevent LLMs from being distracted.

\noindent \textbf{Challenges and open questions:}
Existing pentesting planners can be broadly classified into two categories based on whether they rely on LLMs.
The first category comprises traditional planning methods such as POMDPs and classical planning.
These approaches offer a clear logical structure.
However, they struggle to scale to real-world pentesting environments, where both the state space and action space are large, dynamic, and highly complex.
The second category relies on LLMs to perform planning.
These approaches do not require formal definitions of actions or states, which greatly simplifies system design.
However, the resulting plans often suffer from logical inconsistency and poor long-term coherence.
Moreover, the black-box nature of LLMs makes the planning difficult to control, interpret, or systematically improve.
Therefore, there are two research directions.
The first one is automatically extracting structured knowledge to enable formal planning algorithms in complex penetration scenarios.
The second one is enhancing LLM-based planning so that the generated plans exhibit stronger logical consistency and coherence.

\subsubsection{Executor}
The executor is responsible for (1) translating the planning results into concrete, executable commands and (2) executing those commands on real systems.
Systems without LLMs can only provide commands within a very narrow scope (e.g, a small set of Metasploit exploitations).
Some of them (e.g., \ChainReactor) require human operators to execute the generated attack plans~\cite{depasquale_chainreactor}.
LLMs enable fine-grained generation of commands and code, leading to LLM-driven executors in pentesting, such as the executor of \PentestGPT.
However, these executors cannot interact with target environments, so the generated commands still depend on human operators to execute.
With the rise of LLM-based tool calling, systems like \PentestAgent, \AutoPentester, \CAI are able to synthesize, execute, and iterate commands, significantly reducing human intervention.
Retrieval-augmented generation (RAG) has also been widely adopted in executor.
\PentestAgent, \AutoPentester, \VulnBot, and \PenHeal, for example, combine retrieved code snippets, articles, and previous actions with RAG pipelines to improve the quality of LLM-generated commands.

\noindent \textbf{Challenges and open questions:}
A major challenge for executors is simulating human-like behaviors and interactions.
Many attack vectors only appear through GUIs or interactive workflows where text-only commands and tools are less effective, especially in web penetration scenarios.
Mimicking human behaviors (e.g., mouse and keyboard actions) triggers such attack vectors and helps evade defenses.
Although the Computer-User Interaction Simulation Agent (CUA) has been proposed to mimic human behaviors in interface operations~\cite{wang2025opencua,yang2025gta1}, no existing work has yet applied it for pentesting.
Another challenge is effectively leveraging specialized tools that may lie outside an LLMs' training data.

\subsubsection{Perceptor}
The perceptor is responsible for converting heterogeneous, unstructured data, such as tool outputs and error messages, into representations that the planner can use.
When planning relies solely on LLMs, this unstructured data can be provided directly to the LLM as context, so a dedicated perceptor is unnecessary.
For planners that depend on structured intermediate representations (e.g., PTTs or to-do lists), the perceptor uses an LLM to translate heterogeneous data into the data structures, such as a PTT branch or an item in the to-do list.
For a classical planner, unstructured information is mapped to symbolic predicates, either through manually crafted rules or by an LLM.

\noindent \textbf{Challenges and open questions:}
Existing work focuses primarily on textual data, while visual information is also important in pentesting.
For example, analysts may need to infer a web application’s functionality from its user interface or extract data embedded within images (e.g., reading CAPTCHA).
Developing future perceptors, therefore, requires addressing the challenges of robust visual understanding.
Although an increasing number of LLMs and multimodal models now offer image-analysis capabilities, to the best of our knowledge, no prior work has leveraged visual artifacts effectively in the context of pentesting.


\begin{table*}[ht]
\centering
\caption{Taxonomy of automated penetration testing systems based on the PEP design paradigm.}
\begin{tabular}{l|l|l|l}
\toprule
\textbf{System} & \textbf{Planner} & \textbf{Executor} & \textbf{Perceptor} \\
\midrule
\ChainReactor & Classical Planning & Predefined Actions + Human Operators & Rules + LLM (PDDL predicates) \\
\hline
\PentestGPT & LLM + Penetration Tree & LLM + Human Operators & LLM \\
\hline
AutoPT & LLM + Finite State Machine  & LLM + Agents & LLM\\
\hline
\PentestAgent & LLM + CVE-Exploit Mapping & LLM + RAG (code snippets) + Agents & LLM \\
\hline
AutoAttacker & LLM + Situation Summary & LLM + RAG (previous tasks) + Agents & LLM \\ \hline
\VulnBot & LLM + Penetration Task Graph & LLM + RAG (previous tasks) + Agents & LLM \\
\hline
\PenHeal & LLM + Penetration Tree & LLM + RAG (previous commands) + Agents & LLM\\
\hline
\CAI & LLM & Multiple Tool Agents & -\\
\hline
AutoPentester & LLM + Modified Penetration Tree & LLM + RAG (articles) + Agents & LLM \\
\hline
CheckMate & Classical Planning+ & LLM + Predefined Actions + Agents & LLM \\
\bottomrule
\end{tabular}
\label{tab:pentesting-systems}
\end{table*}


\subsection{Classical Planning}
Classical planners operate on state representations explicitly defined by predicates.
Every action includes clearly defined preconditions and effects, which allow the planner to know exactly which actions are applicable at any given state.
This symbolic grounding guarantees that valid actions are not overlooked, actions are only applied when their preconditions are satisfied, and all changes to the world state are explicitly and consistently tracked throughout the planning process.
In addition, classical planning algorithms guarantee that if a valid sequence of actions exists, the planner must be able to discover it~\cite{blum1997fast}.
Moreover, every intermediate step in the resulting plan is logically consistent with the defined preconditions and effects, ensuring that each causal dependency is correctly maintained.
Consequently, even for long action chains, the planner constructs solutions in a step-by-step manner, preserving the causal structure throughout the entire planning process.

In contrast, LLM-based planning relies on implicit, language-based reasoning~\cite{kambhampati2024llms,cao2025large}.
It lacks a persistent and structured memory of the world state, which makes it prone to forgetting past actions, repeating steps, or hallucinating outcomes, especially as the reasoning chain grows longer~\cite{zhang2025mitigating,ji2024testing,yao2025reasoning}.
This also leads to skipped steps in the action sequence or invalid transitions.
It also suffers from the limited context windows~\cite{liu2024lost} (e.g., 8K–128K tokens), which restrict its ability to retain long-term planning structure~\cite{cao2025large}, especially in complex tasks like penetration testing.

\section{Evaluation of Existing Pentesting Systems}
\subsection{Experimental Methodology \& Setup}

\subsubsection{Benchmark Datasets}
We adopt Vulhub~\cite{vulhub}, a community-maintained collection of containerized vulnerable environments, as the basis of our benchmark.
From this repository, we randomly sampled 120 containers for evaluation.
All target Docker images were anonymized, preventing the evaluation system from recognizing them as Vulhub challenges.
Compared with recent work~\cite{wu2024autopt,deng2024pentestgpt,shen2025pentestagent}, our benchmark is the largest of its kind to date.
We exclude puzzle-like challenges~\cite{picoctf} such as those from HackTheBox~\cite{hackthebox}, which emphasize more on CTF-style tricks.
In addition, these challenge sets (e.g.,~\cite{shao2024nyu}) have extensive public writeups, many of which are likely included in LLM training corpora, posing a risk of data contamination.
To maintain experimental fairness and unbiasedness, we therefore do not incorporate them into our benchmark.

\subsubsection{Metrics}
To measure the actual penetration capability and progress of a pentest, we propose eleven milestones that cover the typical pentesting lifecycle.
Please note that we did not adopt the ``sub-tasks" used in some prior work~\cite{deng2024pentestgpt} as the metric.
Because those sub-tasks emphasize completing specific activities rather than demonstrating meaningful impact.
For example, a sub-task such as ``web enumeration" can be checked off after enumeration was performed; that alone does not show that the tester discovered key information or used it effectively in later phases of the test.
For this reason, we define milestones to assess real progress.
For each pentesting engagement, we manually compare the testing process against the ground truth and judge which milestones have been achieved.
The milestones are as follows:
\begin{itemize}[leftmargin=1em]
    \item M1: Successfully enumerating network hosts, open ports, and running services.
    \item M2: Discovering multiple potential attack vectors (e.g., target services or software) without confirming the actual exploitable one.
    \item M3: Confirming and precisely localizing specific attack vectors susceptible to exploitation.
    \item M4: Obtaining or generating an exploitation command, code, or method.
    \item M5: Successfully executing the exploit that triggers the vulnerability or verifies the PoC.
    \item M6: Successfully executing arbitrary commands on the target system.
    \item M7: Establishing an interactive shell session with user-level privileges.
    \item M8: Discovering a viable privilege escalation method.
    \item M9: Establishing an interactive shell with elevated privileges (root on Linux/Unix, Administrator/SYSTEM on Windows).
    \item M10: Successful lateral movement.
    \item M11: Obtaining authentication credentials or private data in any format.
\end{itemize}

The milestones exhibit both sequential dependencies and parallel paths.
In a typical linear progression, the sequential flow follows the pattern from M1 to M9, with each milestone requiring completion before advancing to the next.
However, M10 and M11 can be pursued in parallel.
For example, the tester may simultaneously explore lateral movement and privilege elevation on compromised systems, and user privacy and credentials can be obtained at any stage throughout the pentesting process.
This milestone framework enables a more nuanced assessment of penetration progress and penetration capabilities.

\subsubsection{Baselines and Evaluation Criteria}
We chose four open-source, well-recognized pentesting systems, \PentestGPT~\cite{deng2024pentestgpt}, \PentestAgent~\cite{shen2025pentestagent}, \CAI~\cite{mayoralvilches2025caiopenbugbountyready}, and AutoPentester~\cite{ginige2025autopentester}, as the baselines.
Related works such as \AutoAttacker~\cite{xu2024autoattacker} and Penheal\cite{huang2023penheal}, as well as commercial systems like XBOW\cite{XBOW}, do not release their code, making independent reproduction infeasible.
\ChainReactor~\cite{depasquale_chainreactor} provides an open-source implementation but narrowly focuses on privilege escalation attacks.
Additionally, while several open-source pentesting toolkits~\cite{hexstrike-ai,PentestAgent,AutoPentestGPT} support automation primitives, they lack autonomous planning modules and thus cannot perform end-to-end automated pentesting.
Lastly, traditional automated pentesting works are excluded because they neither offer reproducible implementations nor have automation capabilities.

\noindent \textbf{Minimal Human Intervention.}
We noticed that some systems still rely on human intervention, such as interpreting system outputs, extracting key information, and guiding LLMs in choosing tools.
We believe rigorous evaluation should minimize human involvement to ensure consistency, fairness, and unbiasedness.
Accordingly, we followed the principle of minimal human intervention.
For systems that could not operate in a fully hands-off-the-keyboard manner, we allowed only essential interactions, strictly limited to selecting default options, executing provided commands, reporting execution outcomes, etc, without offering external knowledge or guidance.

\subsection{Comparative Evaluation of Existing Systems}
In addition to the baselines representing the prior work on automated pentesting, we also evaluated the penetration capabilities of three out-of-the-box LLM agents: Claude Code\cite{anthropic_claude_code} + Sonnet 4.5, Codex\cite{openai_codex} + o4-mini, and Gemini Code Assistant\cite{google_code_assist} + Gemini Pro 2.5.
We fed each system the same initial prompt with the task description and allowed them to use any tool that ships in a standard Kali Linux distribution.
Beyond that, we provided no additional hints or human intervention.
Any single step that stalled for more than two hours was terminated and counted as a failure.
For \PentestGPT\cite{deng2024pentestgpt}, \PentestAgent\cite{PentestAgent}, and \CAI\cite{mayoralvilches2025caiopenbugbountyready}, we employed the most powerful LLM that each system supports.
We measured the percentage of targets that each system advanced to each milestone.

\begin{figure*}
    \centering
    \includegraphics[width=0.8\linewidth]{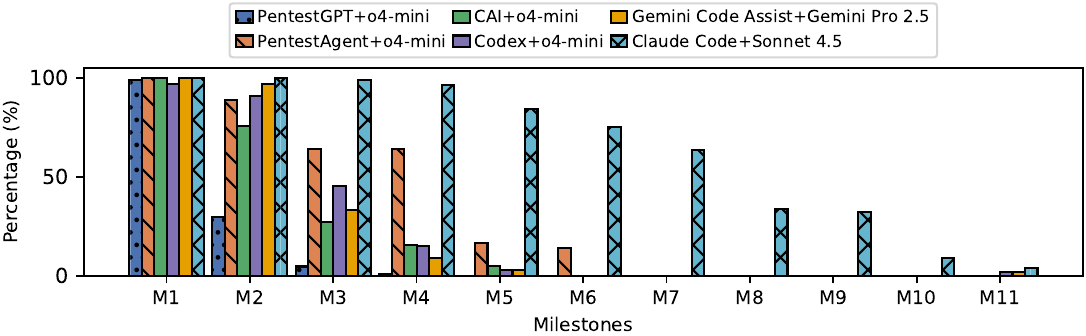}
    \caption{Comparison of penetration capabilities of existing automated pentesting systems on Vulhub benchmark.}
    \label{fig:claude_existing_work_compare}
\end{figure*}

The results shown in Figure~\ref{fig:claude_existing_work_compare} indicate that Claude Code + Sonnet 4.5 consistently outperforms all other systems across almost all milestones.
The performance of \PentestGPT drops sharply after M1, indicating its limited pentesting ability without human intervention.
Although the remaining systems completed the early milestones involving basic reconnaissance and enumeration, their performance diverged significantly once the workflow demands deeper reasoning and planning, as well as exploit development.
\PentestAgent outperforms \CAI, Codex, and Gemini Code Assist due to its online exploit-search strategy, but it still fails to make progress beyond M4.
In contrast, Claude Code maintains strong performance through M7 and still achieves some success in later stages, demonstrating a significantly better capability in multi-step penetration tasks.
Please note that M8 to M11 correspond to lateral movement, privilege escalation, and credential leakage.
Because the Vulhub dataset simulates single–application vulnerabilities, it may not provide the attack paths necessary to reach these milestones.
Overall, these results highlight two key findings.
First, the out-of-the-box Claude Code + Sonnet 4.5 demonstrates substantially stronger capabilities for automated penetration testing than all prior systems evaluated in this domain.
Second, this level of capability is not uniform across LLM-based code agents: Codex and Gemini Code Assist fail to progress beyond basic scanning and enumeration, whereas Claude Code consistently performs a larger number of successful follow-on actions after initial discovery.
A detailed analysis of each system’s penetration process and the factors causing the performance gaps is presented in \S\ref{sec:code-assistant-limitations-and-capabilities}.

\subsection{Discussion on Capabilities and Limitations}\label{sec:code-assistant-limitations-and-capabilities}
To investigate the reasons behind the strong performance of Claude Code + Sonnet 4.5, we conducted an ablation experiment with two modified configurations: (1) replacing Claude Code with an alternative agent, OpenInterpreter, while retaining Sonnet 4.5 as the backend LLM, and (2) retaining Claude Code as the agent but substituting Sonnet 4.5 with GPT-o4-mini.
All other setups were the same.
The results show that both alternative configurations had a substantial drop in performance and failed to achieve M3 on any task.
The pentesting processes also revealed two major weaknesses.
First, the agents occasionally failed to proceed independently and required human input to determine the next step.
Second, they generated redundant steps, such as creating local files or performing unrelated checks before executing actions like port scanning.
Overall, these results suggest that the strong pentesting capability derives from the combination of both Claude Code’s agentic control and Sonnet 4.5’s underlying model capabilities, and that removing either component leads to a significant loss in capability.

We then analyzed the detailed pentesting processes of all systems to answer two key questions: how does Claude Code outperform other systems, and where does it still fall short?
\subsubsection{Advantages of Claude Code}
\PentestGPT relies on a human-in-the-loop workflow, while other systems incorporate command-execution capabilities, enabling a higher degree of automation.
The LLM code agents are better at iteratively debugging and modifying commands based on the execution results compared to \CAI.
Although \PentestAgent excels in searching and leveraging online exploits, its performance lags in non-exploit tasks, such as enumeration and application probing.
Among those LLM agents, Claude Code stands out with several strengths.
First, Codex and Gemini Code Assist discover narrower attack surfaces and often select slower and less effective approaches (e.g., excessive password brute-forcing or path enumeration), while Claude Code can discover broader attack surfaces.
Second, Codex and Gemini Code Assist demonstrate limited self-reflection and adjustment capabilities.
They often fail to recognize when a chosen route is unproductive and may remain stalled on the same command for extended periods.
In contrast, Claude Code continuously monitors command execution, detects potential deadlocks, and autonomously terminates stalled processes to pursue alternative actions.
Claude Code is even capable of parallel multitasking.
It can start a new task while another is still running, and if the new task produces more valuable findings, it will automatically kill the previous ones.
Lastly, despite being explicitly instructed not to request human input, Codex and Gemini Code Assist frequently paused to request user decisions or inputs.

\subsubsection{Limitations of Claude Code}
Despite its leading performance, we identify three limitations of Claude Code, which also exist in other LLM agents.

\noindent \textbf{Claude Code often fails to maintain a coherent attack plan:}
Claude Code does not follow a consistent, strategic sequence of actions, leading to repeated work, abandoned partial attempts, and unstable performance.
Its decision-making tends to execute whatever actions come to “mind” first.
For example, after identifying vulnerable applications, the agent may initially search for exploits in Metasploit, then abruptly switch to GitHub.
Even after downloading a potential exploit script, it may suddenly abandon that path to write a new exploit for a different target or pursue an unrelated attack vector.
This lack of strategic consistency also appears in basic tasks such as port scanning.
Claude Code may perform a full port scan in one session, scan only the top 1000 ports in another, or scan its own list of “common ports” in another task.
Such unpredictability makes the pentesting workflow difficult to anticipate or control.
It causes the agent to diverge from optimal methodologies, overlook viable attack vectors, and waste significant time.

\noindent \textbf{Claude Code struggles with long-term and experience-driven reasoning:}
Reasoning in pentesting means inferring causal relationships from received information to feasible actions.
For example, when a web application is discovered, a typical reasoning process would involve enumerating version information, identifying all relevant exploits, and then selecting the most promising one to attempt.
However, this type of long-term, multi-step reasoning is difficult for LLMs.
They may skip enumeration or investigation steps and jump directly to generating an exploit based on their internal knowledge.
While this can occasionally speed up the process, it also increases the risks of hallucinated steps, inconsistent performance, and unnecessary token usage.
In addition, LLMs struggle with experience-driven reasoning, the ability to leverage subtle cues in pentesting.
For instance, a URL pattern like “/node/\{number\}” can hint at a Drupal backend.
An experienced human pentester would immediately consider Drupal-specific attack paths after identifying such URLs.
In contrast, LLMs often fail to recognize this kind of implicit linkage, leading to missed opportunities.

\noindent \textbf{Claude Code has difficulty using specialized pentesting tools:}
We also observed a tendency that Claude Code favors crafting custom scripts instead of first leveraging established, specialized pentesting tools, which diverges from human pentesting methodology.
For example, Claude Code frequently generates custom curl commands to probe web applications, even though thousands of highly effective Nuclei scanning templates already exist and would provide broader, faster, and more reliable coverage.
We assume this is largely because such tools appear less frequently in LLMs' training data.

\section{System Design}
\subsection{Overview}
\begin{figure}
    \centering
    \includegraphics[width=\linewidth]{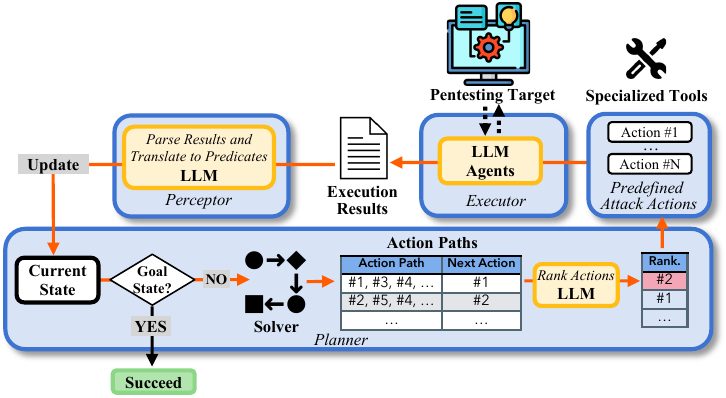}
    \caption{Overview of \SysName. The orange arrow shows the iterative loop of classical planning+. The current state is initialized before the planning starts.}
    \label{fig:flowchart}
\end{figure}

In this paper, we present \SysName, a system designed to overcome limitations of existing LLM-based pentesting frameworks.
Following the PEP diagram proposed in \S\ref{sec:pep-model-and-related-work}, \SysName consists of three major components: classical planning+ as the planner, an LLM agent as the executor, and an LLM as the perceptor.
The overall design of \SysName is illustrated in Figure~\ref{fig:flowchart}.
Specifically, we introduced predefined attack actions to expand LLM’s knowledge on the specialized tools.
Classical planning+ is leveraged to plan the next action, which is executed by an LLM agent.
An LLM is used to interpret execution results and update the planner for further planning.
Instead of relying on the LLM agent for the entire pentesting workflow, \SysName restricts the LLM’s role to a pure perceptor and a simple-task executor.
This design leverages the LLM agent’s strong executing and interpreting capabilities while relieving it of long-horizon planning and reasoning, which are handled by the classical planner.

\subsection{Predefined Attack Actions}
As previously mentioned, existing general-purpose LLM agents lack knowledge of specialized tools during pentesting.
To address this, we introduce predefined attack actions to expand their knowledge base.
We explicitly predefine niche and fine-grained tools such as Metasploit modules, NSE scripts, and Nuclei templates as ``actions", which are considered by the planner.
Predefined attack actions also help avoid the inconsistency and errors in LLM-command-generation.
In pentesting, most commands adhere to a consistent structure.
For example, when executing a default port scan using \texttt{nmap -Pn -sC -sV -p- -oN - \#\{target\}}, the command structure remains largely consistent, while the only part that usually changes is \texttt{\#\{target\}}.
However, the next-token prediction mechanism of LLMs is increasingly unstable, and error-prone when generating long commands.
In contrast, predefined attack actions provide the core structure and options of the command, leaving only parameters like \texttt{\#\{target\}} to be specified, significantly reducing the risk of generating incorrect commands.

Predefined attack action offers an alternative approach to expanding an LLM’s knowledge base without relying on traditional RAG or fine-tuning.
Fine-tuning LLMs is often costly, time-consuming, and difficult to scale.
RAG, while flexible, depends on retrieving document snippets and the model’s ability to interpret those snippets and synthesize commands.
In contrast, predefined attack actions offer explicit, well-structured, executable commands.
By defining preconditions of actions (we will introduce this later), these actions can be retrieved more accurately, efficiently, and interpretably than relying solely on embedding-based similarity search in RAG.

\subsection{Planner}
We propose classical planning+ as the planner of \SysName, which encodes the causal relationships explicitly and maintains a persistent and coherent plan throughout the pentesting.
\subsubsection{Causal Relationships}
Causal relationships are encoded explicitly through the preconditions and effects associated with each action.
For example, once a web enumeration action identifies a specific web application, a pentester would naturally consider all relevant Metasploit modules, NSE scripts, and Nuclei templates associated with that application.
In classical planning, the discovered web application is an effect of the enumeration action and simultaneously serves as a precondition of those subsequent actions.
The set of factors used as preconditions and effects is flexible and can be customized or extended as needed.
In our current design, these factors include elements such as the identified application, CVEs, URLs, usernames, passwords, etc. 
By encoding these causal dependencies directly, the system gains stronger built-in causal reasoning capabilities and reduces the need for the LLM to perform complex long-horizon reasoning on its own.

\subsubsection{Classical Planning+}\label{sec:dynamic-pddl-updating}
Classical planning+ is proposed to address the limitations of traditional classical planning in dynamic, non-deterministic, and partially observable tasks.

\noindent \textbf{Non-Deterministic Action Effects:} 
Pentesting inherently involves uncertainty and incomplete information.
For instance, the result of a port scan is not known until finished, and the outcome of an exploit attempt is often unpredictable until it is executed.
However, traditional classical planning assumes a static, deterministic, and fully observable target, where all action effects are determined, and the state of the target is completely specified before the planning starts.
Some pentesting systems use complex models to encode uncertainty, which are difficult to scale in the real world.
In \SysName, we propose classical planning+, leveraging LLMs to dynamically determine action effects.
Since it updates action effects at runtime, complete knowledge is no longer required before planning begins.
Specifically, we define the non-deterministic effect to indicate that the effect of an action is unknown until it is executed.
Once an action with a non-deterministic effect is executed, LLMs are invoked to analyze the execution outcome and generate concrete effect predicates.
We describe this process in \S\ref{sec:dynamic-pddl-updating}, along with a concrete example.
Through this mechanism, we successfully extend classical planning to dynamic, non-deterministic, partially-observable scenarios.

\begin{figure*}[h]
    \centering
    \includegraphics[width=0.78\linewidth]{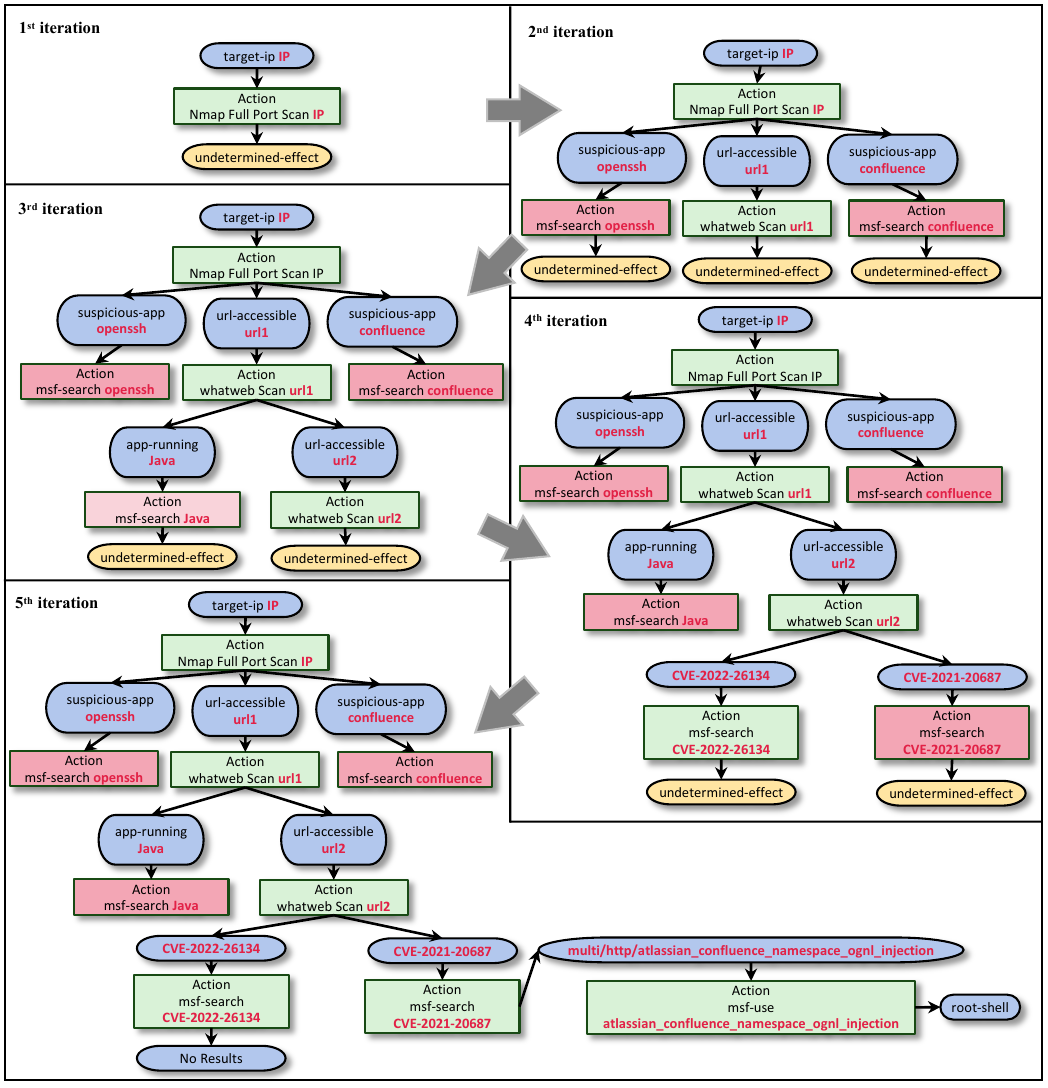}
    \caption{A pentesting workflow driven by classical planning+. Each panel shows one planning-execution-perception iteration. Blue rounded ovals are predicates that link actions across iterations; yellow rounded ovals denote non-deterministic action effects. Rectangular boxes list feasible actions available during the engagement, and light-green rectangles indicate the actions chosen by the planner for execution in that iteration. Arrows show how actions are connected with predicates.}
    \label{fig:penetration_example}
\end{figure*}

Classical planning+ begins from the initial state representing all prior knowledge about the target.
In each iteration, the planner checks whether the goal is reachable under the current state.
If it is, the planner executes the action sequence leading to the goal.
If not, the planner produces a list of applicable actions by checking the preconditions of each action.
Next, the LLM executes the optimal action from this list based on its knowledge and updates the initial state.
If the executed action has a non-deterministic effect, the LLM is invoked to analyze the execution output and translate it into concrete predicates.
This process is repeated iteratively until either the goal is met or all possible actions have been explored.
Compared to LLM agents, classical planning+ provides a more structured planning engine by presenting a plan as a directed acyclic graph. 
It offers several advantages in pentesting planning.
First, it exhaustively explores the entire action space, avoiding missing available actions, especially in scenarios with a large number of actions or long action sequences.
Second, it avoids repeating previously executed actions or jumping across different directions, which is a common failure in LLM-based planning.
Moreover, the planning process is both visible and interpretable.
We illustrate classical planning+ using an example in Figure~\ref{fig:penetration_example}.

\begin{algorithm}
\caption{Iterative Planning for Penetration Testing under Partial Knowledge}
\begin{algorithmic}[1]
\State \textbf{Input:} Domain $D$ with predefined actions, initial knowledge $I_0$
\State \textbf{Initialize:} $S \gets I_0$ \Comment{Initial state from known information (e.g., IP)}
\While{termination condition not met \textbf{and} actions remain}
    \State applicableActions $\gets$ \{\}
    \ForAll{action $a$ in domain $D$}
        \If{$a$ is reachable from state $S$}
            \State seq $\gets$ plan($S$, $a$)
            \State applicableActions.add(seq.first())
        \EndIf
    \EndFor
    \State nextAction $\gets$ LLM\_Select(applicableActions)
    \State result $\gets$ Execute(nextAction)
    \If{nextAction has deterministic effects}
        \State $S \gets S\ \cup$ effects(nextAction)
    \Else
        \State preds $\gets$ Parse\_NonDeterministic\_Effects(result)
        \State $S \gets S\ \cup$ preds
    \EndIf
\EndWhile
\If{goal is not achieved}
    \State Report failure: challenge unsolvable.
\EndIf
\end{algorithmic}
\end{algorithm}

\subsection{Executor}
Once the next action is determined, the system should execute it without human intervention.
To do so, it must select the appropriate tools, reliably generate precise, executable instructions, and configure all required parameters.
Given the strong execution abilities of LLM-based code agents, \SysName employs an LLM agent as its executor.
Each predefined action is paired with a concise, action-specific prompt that guides the agent.
These prompts specify the required tool and command structure, along with placeholders for parameters.
For instance, for a network-scanning action, the prompt clearly outlines the expected flags and arguments, while still allowing task-specific values (the IP address) to be injected.
These placeholders are automatically populated by the classical planner, ensuring that critical parameters, such as a module name for an exploitation step, are determined deterministically rather than by the LLM, thereby reducing the risk of hallucinations.
After the planner selects an action, \SysName provides the corresponding prompt to the LLM executor, which performs the command and returns the resulting output for downstream processing.

\subsection{Perceptor}
The perceptor bridges the executor and the planner: it analyzes the execution results in heterogeneous formats and content, translates them into the representation that the planner can use for subsequent planning.
In \SysName, the perceptor translates the outputs into predicates defined in classical planning+, which are then used to update the current state.
\SysName has two types of perceptors: a rule-based perceptor and an LLM-based perceptor.
The rule-based perceptor parses structured outputs and maps them to the corresponding predicates, avoiding the randomness introduced when using LLMs.
For example, the JSON result returned from a Metasploit search can be directly mapped to a predicate \textit{(msf-module-available ?exploit-name)} for simplicity.
The LLM-based perceptor leverages LLMs to interpret unstructured outputs and produce predicates defined in classical planning+.

\section{Evaluation}
\subsection{Penetration Capability}
\begin{figure}
    \centering
    \includegraphics[width=0.9\linewidth]{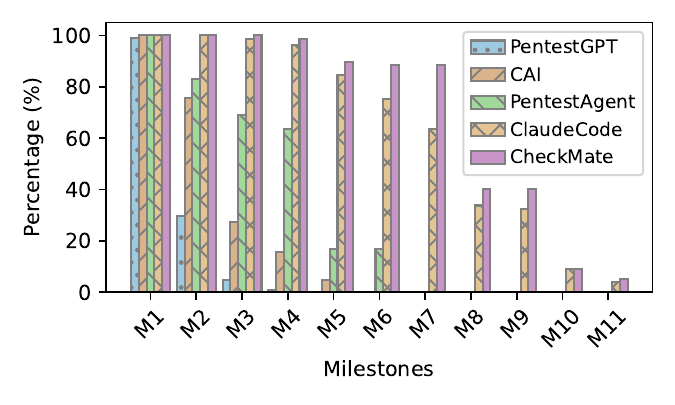}
    \caption{Comparison of Claude Code with \SysName on Vulhub benchmark}
    \label{fig:claude_checkmate_compare}
\end{figure}

We first evaluate the penetration capability of \SysName, compared to existing work.
We adopt the same benchmark dataset, metrics, and experimental setup as described in Section 3.
The results are shown in Figure~\ref{fig:claude_checkmate_compare}.
\SysName demonstrates substantially stronger penetration capability than all baselines, as evidenced by its progress across all milestones.
Notably, 88\% of its penetration attempts reach milestone M7, whereas prior work, except Claude Code, rarely progresses beyond M4.
Furthermore, \SysName shows advantages over Claude Code at the higher milestones, particularly in the success rates for M6 and M7, indicating improved effectiveness in executing exploits and successfully obtaining a shell.
These gains result from the explicitly defined, fine-grained actions and a structured planning strategy.
By planning all available actions before committing to any specific attack path, \SysName avoids becoming trapped in unproductive branches and maintains steady progress toward deeper system compromise.

\subsection{Efficiency}\label{sec:efficiency-evaluation}
\begin{figure}[htbp]
    \centering
    \begin{subfigure}{0.9\linewidth}
        \centering
        \includegraphics[width=\linewidth]{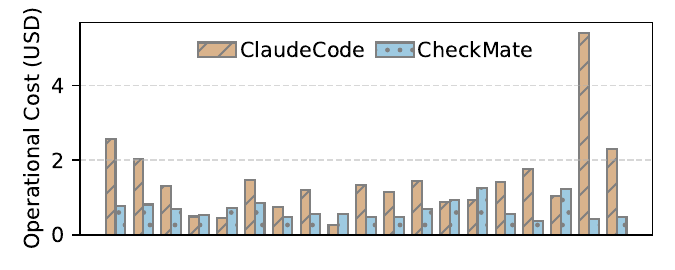}
        \caption{Monetary costs.}
        \label{fig:top}
    \end{subfigure}
    

    \begin{subfigure}{0.9\linewidth}
        \centering
        \includegraphics[width=\linewidth]{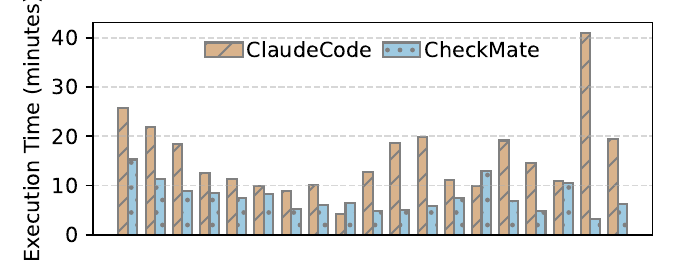}
        \caption{Time consumed.}
        \label{fig:bottom}
    \end{subfigure}

    \caption{Efficiency comparison between ClaudeCode and \SysName.}
    \label{fig:cost_and_time_comparison}
\end{figure}

In this section, we evaluate both the efficiency and cost of \SysName.
We selected 20 penetration tasks that \SysName and Claude Code were both able to successfully complete.
Under the same LLM model setting, we compared the total monetary cost, representing the amount of LLM tokens consumed, and the time required to finish each task.
The results are summarized in Figure~\ref{fig:cost_and_time_comparison}.
On average, CheckMate has a total cost of \$0.68, which is 53\% lower than that of Claude Code under identical conditions.
This reduction in token consumption can be attributed to the use of classical planning for strategy formulation.
In contrast, Claude Code relies entirely on text-based reasoning, where every intermediate thought and plan must be expressed in natural language, leading to substantial token overhead.
By adopting a symbolic and formalized planning mechanism, \SysName avoids using the LLM to “generate” its reasoning process, thereby concentrating the model’s generation capacity on executing actions and interpreting outputs.
The average time consumed for \SysName is 7.75 minutes, which is 54\% lower than Claude Code.

\subsection{Stability}
In this section, we evaluate the stability of the pentesting process, i.e., whether the system demonstrates consistent performance across repeated executions of the same task.
To assess this, we execute each task three times and record the results, costs, and time consumption for each run.
We then compute the success rate (i.e., the proportion of runs in which all three attempts successfully achieve penetration) and the \textit{Coefficient of Variation} (a scale-independent measure of dispersion) of both cost and time.
The aggregated results are summarized in Table~\ref{table:stability-comparison}.
About 25\% of the tasks cannot be solved consistently by Claude Code across all three attempts. In addition, \SysName demonstrates higher consistency in both LLM token usage and execution time.
These improvements stem from \SysName’s adoption of a more structured planning engine, which reduces unnecessary fluctuations introduced by the LLM.

\begin{table}[h!]\footnotesize
\centering
\caption{Stability comparison between \SysName and Claude Code.}
\begin{tabular}{c|c|c}
\toprule
                 & \SysName &  Claude Code \\ \midrule
                 
Success Rate for all Attempts ($\uparrow$)        &   100\%   &   75\%  \\ 
Coefficient of Variation - Cost ($\downarrow$)         &   0.129    &  0.451   \\ 
Coefficient of Variation - Time ($\downarrow$)         &  0.093   &   0.325  \\
\bottomrule
\end{tabular}
\label{table:stability-comparison}
\end{table}

\subsection{Case Study}
\begin{figure}
    \centering
    \includegraphics[width=0.85\linewidth]{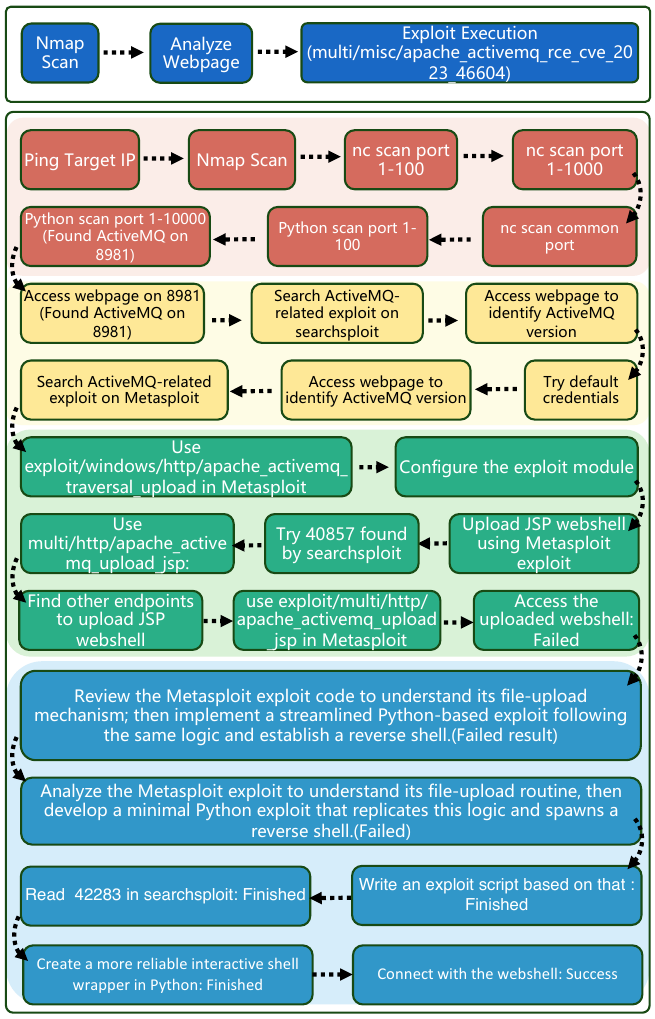}
    \caption{Top box: \SysName’s workflow. Bottom box: Claude Code’s workflow. Colors show stages. pink: reconnaissance, yellow: search/analysis, green: Metasploit/SearchSploit exploitation (failed), blue: autonomous exploitation.}
    \label{fig:placeholder}
\end{figure}

We analyze a specific example in detail to illustrate differences during pentesting between \SysName and Claude Code.
In this case, \SysName completed the penetration in only three steps; Claude Code, by contrast, used 26 steps, many of which were added because of redundancy, premature abandonment, distractions, and incomplete planning and reasoning.
The target is an old version of Apache ActiveMQ (an open-source messaging middleware that supports Java messaging services, clustering, and the Spring framework) from Vulhub.
\SysName began with a full-port Nmap scan plus fingerprinting and script probes.
It discovered two open ports (22 and 8191), identified that Apache ActiveMQ was running, and associated that service with likely CVEs and corresponding Metasploit modules.
Rather than rushing straight to exploitation, \SysName chose, from the feasible action set, to analyze the web interface to further confirm the ActiveMQ version.
That analysis verified that an ActiveMQ Console was running and revealed the precise version, 5.11.1.
Armed with that version information, in the third step, \SysName selected Metasploit’s ``multi/misc/apache\_activemq\_rce\_cve\_2023\_46604" module and ran the exploit, ultimately obtaining a root shell on the target.
At each stage, the LLM evaluated all available actions and prioritized them, demonstrating a highly planned and systematic approach.

By contrast, Claude Code’s test was a largely ad-hoc process, showing exploratory and blind trial-and-error behavior.
Claude Code first tried to ping the target IP and run nmap scans, but abandoned both because of insufficient permissions.
In fact, both commands failed due to missing socket privileges and would have succeeded with ``sudo".
Instead of modifying the commands to add the necessary privileges, Claude Code pivoted to using Netcat and writing Python scripts, which are more complex methods for port scanning.
Its port-scanning also lacked a coherent plan: it scanned the first 100 ports, then 1,000 ports, then ``common" ports, and only later broadened the range, thus finally finding port 8191 open.
It is a risky strategy because when Claude Code hit a rabbit hole on common ports, it tended to pursue the wrong path, wasting time or risking failure.
Claude Code also struggled to remain focused on a single attack path.
While attempting to determine the ActiveMQ version, it would abruptly switch to trying the default-credential brute force.
After selecting and spending a long time configuring a Metasploit module, it might suddenly divert to investigating another script found on Exploit-DB, creating needless context switches and time loss.
Finally, because Claude Code lacked explicit, structured reasoning, it failed to map the discovered ActiveMQ version to the most appropriate CVE.
As a result, it missed the more effective Metasploit module and wasted excessive time on two suboptimal exploits.

\subsection{Ablation Study}
\begin{figure}
    \centering
    \includegraphics[width=0.85\linewidth]{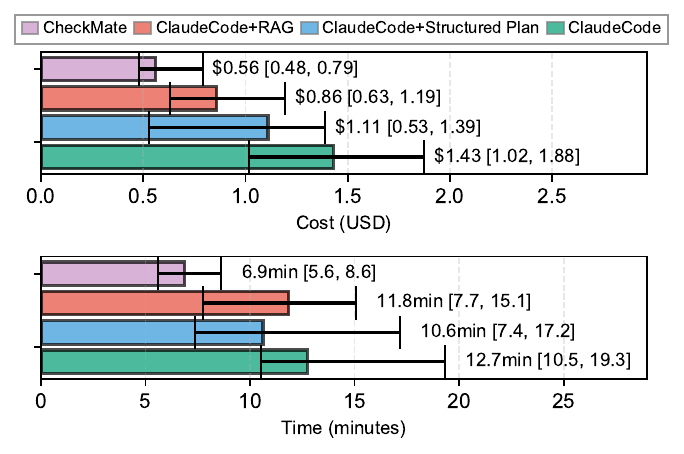}
    \caption{Cost and time comparison. (a) Median API costs in USD. (b) Median execution time in minutes. Error bars represent the interquartile range (25th-75th percentile).}
    \label{fig:ablation}
\end{figure}

In this section, we conduct an ablation study by comparing \SysName with two commonly used strategies for enhancing LLM-based systems.
First, we compare \SysName against the RAG-based approach, which is an alternative strategy for expanding an LLM’s knowledge base.
We embedded metadata of specialized penetration tools, including more than 14 thousand Metasploit modules, NSE scripts, and Nuclei templates, as the document database and implemented a RAG pipeline.
We aim to evaluate whether LLM agents can effectively use external knowledge to improve their penetration capabilities without relying on predefined actions and classical planning+.
Second, we let Claude Code maintain a structured planning file in JSON format rather than using its default to-do list.
The prompting was modified so that after each command execution, Claude Code updates this structured planning file and infers the next step based on the revised state.
This approach reflects common methodologies in prior work that employ structured planning representations to improve an LLM agent’s planning consistency.
For each method, we evaluated the performance on 20 tasks, running each task three times.
All four methods successfully obtained a remote shell at least once.
As shown in Figure~\ref{fig:ablation}, \SysName achieves the lowest overall cost and the shortest execution time, while also delivering the most consistent and efficient performance across test cases.
These results indicate that although incorporating RAG or structured planning files can enhance the efficiency of LLM-based agents, the classical planning+ approach provides the most substantial gains in both efficiency and consistency.

\section{Discussion and Future Work}
\subsection{Actions and States in Pentesting}
Existing work leaves two fundamental questions unanswered: (1) What actions and skills does pentesting require? and (2) How should we represent the state of the target system?
The difficulty arises from the open-ended nature of pentesting.
Unlike tasks with well-defined action and state spaces, pentesting spans the full breadth of a system’s architecture, configurations, vulnerabilities, and defenses, and demands a wide and adaptable skill set.
Current approaches either define fixed, finite sets of skills and states, or depend heavily on black-box LLMs to infer target states and propose actions.
The fixed schemas are too restrictive, while relying on opaque LLMs makes it difficult to systematically improve penetration capabilities.
This gap highlights the need for future work on representing, organizing, extracting, and operationalizing the fragmented knowledge on actions and states in pentesting.

\subsection{Multimodal and UI-Aware Pentesting}
Existing pentesting systems struggle in scenarios that require rich human-computer interaction, as traditional LLM agents are not good at interpreting non-textual information and operating web user interfaces (UIs) like a human.
As a result, tasks that involve understanding visual elements or manipulating dynamic, interactive web components still depend heavily on humans.
Recent advances in multimodal learning and Customizable UI Automation (CUA) offer promising avenues for addressing these limitations~\cite{wang2025opencua,yang2025gta1}, opening up new possibilities for pentesting in complex UI environments.

\section{Conclusion}
In this paper, we first presented a systematic review of existing automated pentesting work through the lens of our Planner-Executor-Perceptor (PEP) paradigm.
Our evaluation shows that the out-of-the-box Claude Code+Sonnet 4.5 substantially outperforms all prior systems in this area.
However, further analysis revealed three limitations of Claude Code.
We thus proposed \SysName, a framework that couples classical planning+ with LLM agents to address these limitations.
Experimental evaluations demonstrated that \SysName outperforms existing systems in penetration capability, efficiency, and stability.

\section*{Ethical Considerations}
This paper presents a practical study on using LLM Agents for pentesting.
All techniques and systems involved are publicly accessible; we have not developed any new zero-day attacks.
All experiments were conducted within authorized virtual environments.
We will contact the service providers to inform them of the potential for their products in offensive scenarios.
This work is intended solely for research and educational purposes, and we do not encourage or endorse any misuse of the discussed techniques.

\bibliographystyle{IEEEtran}
\bibliography{refs}

@article{liu2024lost,
  title={Lost in the middle: How language models use long contexts},
  author={Liu, Nelson F and Lin, Kevin and Hewitt, John and Paranjape, Ashwin and Bevilacqua, Michele and Petroni, Fabio and Liang, Percy},
  journal={Transactions of the Association for Computational Linguistics},
  volume={12},
  pages={157--173},
  year={2024}
}

@article{yao2025reasoning,
  title={Are Reasoning Models More Prone to Hallucination?},
  author={Yao, Zijun and Liu, Yantao and Chen, Yanxu and Chen, Jianhui and Fang, Junfeng and Hou, Lei and Li, Juanzi and Chua, Tat-Seng},
  journal={arXiv preprint arXiv:2505.23646},
  year={2025}
}

@article{zhang2025mitigating,
  title={Mitigating spatial hallucination in large language models for path planning via prompt engineering},
  author={Zhang, Hongjie and Deng, Hourui and Ou, Jie and Feng, Chaosheng},
  journal={Scientific Reports},
  volume={15},
  number={1},
  pages={8881},
  year={2025},
  publisher={Nature Publishing Group UK London}
}

@article{kambhampati2024llms,
  title={Llms can't plan, but can help planning in llm-modulo frameworks},
  author={Kambhampati, Subbarao and Valmeekam, Karthik and Guan, Lin and Verma, Mudit and Stechly, Kaya and Bhambri, Siddhant and Saldyt, Lucas and Murthy, Anil},
  journal={arXiv preprint arXiv:2402.01817},
  year={2024}
}

@book{ghallab2004automated,
  title={Automated Planning: theory and practice},
  author={Ghallab, Malik and Nau, Dana and Traverso, Paolo},
  year={2004},
  publisher={Elsevier}
}

@article{rahman2025llm,
  title={Llm-based data science agents: A survey of capabilities, challenges, and future directions},
  author={Rahman, Mizanur and Bhuiyan, Amran and Islam, Mohammed Saidul and Laskar, Md Tahmid Rahman and Mahbub, Ridwan and Masry, Ahmed and Joty, Shafiq and Hoque, Enamul},
  journal={arXiv preprint arXiv:2510.04023},
  year={2025}
}

@inproceedings{ullah2024llms,
  title={Llms cannot reliably identify and reason about security vulnerabilities (yet?): A comprehensive evaluation, framework, and benchmarks},
  author={Ullah, Saad and Han, Mingji and Pujar, Saurabh and Pearce, Hammond and Coskun, Ayse and Stringhini, Gianluca},
  booktitle={2024 IEEE symposium on security and privacy (SP)},
  pages={862--880},
  year={2024},
  organization={IEEE}
}

@article{guo2025repoaudit,
  title={Repoaudit: An autonomous llm-agent for repository-level code auditing},
  author={Guo, Jinyao and Wang, Chengpeng and Xu, Xiangzhe and Su, Zian and Zhang, Xiangyu},
  journal={arXiv preprint arXiv:2501.18160},
  year={2025}
}

@article{liu2024large,
  title={Large language model-based agents for software engineering: A survey},
  author={Liu, Junwei and Wang, Kaixin and Chen, Yixuan and Peng, Xin and Chen, Zhenpeng and Zhang, Lingming and Lou, Yiling},
  journal={arXiv preprint arXiv:2409.02977},
  year={2024}
}

@article{jin2024llms,
  title={From llms to llm-based agents for software engineering: A survey of current, challenges and future},
  author={Jin, Haolin and Huang, Linghan and Cai, Haipeng and Yan, Jun and Li, Bo and Chen, Huaming},
  journal={arXiv preprint arXiv:2408.02479},
  year={2024}
}

@article{wang2025agents,
  title={Agents in software engineering: Survey, landscape, and vision},
  author={Wang, Yanlin and Zhong, Wanjun and Huang, Yanxian and Shi, Ensheng and Yang, Min and Chen, Jiachi and Li, Hui and Ma, Yuchi and Wang, Qianxiang and Zheng, Zibin},
  journal={Automated Software Engineering},
  volume={32},
  number={2},
  pages={1--36},
  year={2025},
  publisher={Springer}
}

@article{wang2025opencua,
  title={Opencua: Open foundations for computer-use agents},
  author={Wang, Xinyuan and Wang, Bowen and Lu, Dunjie and Yang, Junlin and Xie, Tianbao and Wang, Junli and Deng, Jiaqi and Guo, Xiaole and Xu, Yiheng and Wu, Chen Henry and others},
  journal={arXiv preprint arXiv:2508.09123},
  year={2025}
}

@article{yang2025gta1,
  title={Gta1: Gui test-time scaling agent},
  author={Yang, Yan and Li, Dongxu and Dai, Yutong and Yang, Yuhao and Luo, Ziyang and Zhao, Zirui and Hu, Zhiyuan and Huang, Junzhe and Saha, Amrita and Chen, Zeyuan and others},
  journal={arXiv preprint arXiv:2507.05791},
  year={2025}
}

@misc{hackthebox,
  key          = {HackTheBox},
  title        = {{Hack The Box}},
  howpublished = {\url{https://www.hackthebox.com/}},
  note         = {Accessed: 2025-12-04}
}

@misc{vulhub,
  key          = {Vulhub},
  title        = {{Vulhub}: Open-Source Vulnerable Docker Environments},
  howpublished = {\url{https://vulhub.org/}},
  note         = {Accessed: 2025-12-04}
}

@misc{picoctf,
  key          = {picoCTF},
  title        = {{picoCTF}},
  howpublished = {\url{https://picoctf.org/}},
  note         = {Accessed: 2025-12-04}
}

@article{wu2024autopt,
  title={Autopt: How far are we from the end2end automated web penetration testing?},
  author={Wu, Benlong and Chen, Guoqiang and Chen, Kejiang and Shang, Xiuwei and Han, Jiapeng and He, Yanru and Zhang, Weiming and Yu, Nenghai},
  journal={arXiv preprint arXiv:2411.01236},
  year={2024}
}

@article{kong2025vulnbot,
  title={Vulnbot: Autonomous penetration testing for a multi-agent collaborative framework},
  author={Kong, He and Hu, Die and Ge, Jingguo and Li, Liangxiong and Li, Tong and Wu, Bingzhen},
  journal={arXiv preprint arXiv:2501.13411},
  year={2025}
}

@article{shao2024nyu,
  title={Nyu ctf bench: A scalable open-source benchmark dataset for evaluating llms in offensive security},
  author={Shao, Minghao and Jancheska, Sofija and Udeshi, Meet and Dolan-Gavitt, Brendan and Milner, Kimberly and Chen, Boyuan and Yin, Max and Garg, Siddharth and Krishnamurthy, Prashanth and Khorrami, Farshad and others},
  journal={Advances in Neural Information Processing Systems},
  volume={37},
  pages={57472--57498},
  year={2024}
}

@article{wang2021automatic,
  title={An Automatic Planning-Based Attack Path Discovery Approach from IT to OT Networks},
  author={Wang, Zibo and Zhang, Yaofang and Liu, Zhiyao and Wei, Xiaojie and Chen, Yilu and Wang, Bailing},
  journal={Security and Communication Networks},
  volume={2021},
  number={1},
  pages={1444182},
  year={2021},
  publisher={Wiley Online Library}
}

@article{chen2024survey,
  title={A survey on penetration path planning in automated penetration testing},
  author={Chen, Ziyang and Kang, Fei and Xiong, Xiaobing and Shu, Hui},
  journal={Applied Sciences},
  volume={14},
  number={18},
  pages={8355},
  year={2024},
  publisher={MDPI}
}

@article{obes2013attack,
  title={Attack planning in the real world},
  author={Obes, Jorge Lucangeli and Sarraute, Carlos and Richarte, Gerardo},
  journal={arXiv preprint arXiv:1306.4044},
  year={2013}
}

@inproceedings{hoffmann2015simulated,
  title={Simulated penetration testing: from" Dijkstra" to" Turing Test++"},
  author={Hoffmann, J{\"o}rg},
  booktitle={Proceedings of the international conference on automated planning and scheduling},
  volume={25},
  pages={364--372},
  year={2015}
}

@article{zhou2021autonomous,
  title={Autonomous penetration testing based on improved deep q-network},
  author={Zhou, Shicheng and Liu, Jingju and Hou, Dongdong and Zhong, Xiaofeng and Zhang, Yue},
  journal={Applied Sciences},
  volume={11},
  number={19},
  pages={8823},
  year={2021},
  publisher={MDPI}
}

@article{ghanem2023hierarchical,
  title={Hierarchical reinforcement learning for efficient and effective automated penetration testing of large networks},
  author={Ghanem, Mohamed C and Chen, Thomas M and Nepomuceno, Erivelton G},
  journal={Journal of Intelligent Information Systems},
  volume={60},
  number={2},
  pages={281--303},
  year={2023},
  publisher={Springer}
}

@article{schwartz2019autonomous,
  title={Autonomous penetration testing using reinforcement learning},
  author={Schwartz, Jonathon and Kurniawati, Hanna},
  journal={arXiv preprint arXiv:1905.05965},
  year={2019}
}

@article{sarraute2013penetration,
  title={Penetration testing== POMDP solving?},
  author={Sarraute, Carlos and Buffet, Olivier and Hoffmann, J{\"o}rg},
  journal={arXiv preprint arXiv:1306.4714},
  year={2013}
}

@inproceedings{schwartz2020pomdp+,
  title={Pomdp+ information-decay: Incorporating defender's behaviour in autonomous penetration testing},
  author={Schwartz, Jonathon and Kurniawati, Hanna and El-Mahassni, Edwin},
  booktitle={Proceedings of the International Conference on Automated Planning and Scheduling},
  volume={30},
  pages={235--243},
  year={2020}
}

@inproceedings{sarraute2012pomdps,
  title={POMDPs make better hackers: Accounting for uncertainty in penetration testing},
  author={Sarraute, Carlos and Buffet, Olivier and Hoffmann, J{\"o}rg},
  booktitle={Proceedings of the AAAI Conference on Artificial Intelligence},
  volume={26},
  number={1},
  pages={1816--1824},
  year={2012}
}

@misc{anthropic_claude_code,
  author       = {Anthropic},
  title        = {Claude Code},
  year         = {2025},
  howpublished = {\url{https://www.claude.com/product/claude-code}},
  note         = {Accessed: 2025-11-07}
}

@misc{openai_codex,
  author       = {OpenAI},
  title        = {OpenAI Codex},
  year         = {2025},
  howpublished = {\url{https://openai.com/codex/}},
  note         = {Accessed: 2025-11-07}
}

@misc{google_code_assist,
  author       = {Google},
  title        = {Code Assist},
  year         = {2025},
  howpublished = {\url{https://codeassist.google/}},
  note         = {Accessed: 2025-11-07}
}

@misc{hexstrike-ai,
  author       = {0x4m4},
  title        = {HexStrike AI MCP Agents},
  howpublished = {\url{https://github.com/0x4m4/hexstrike-ai}},
  note         = {Accessed: 2025-10-16},
  year         = {2025}
}

@misc{PentestAgent,
  author       = {GH05TCREW},
  title        = {PentestAgent: All‑in‑one offensive security toolbox with AI agent and MCP architecture},
  howpublished = {\url{https://github.com/GH05TCREW/PentestAgent}},
  note         = {Accessed: 2025-10-16},
  year         = {2025}
}

@misc{AutoPentestGPT,
  author       = {Armur‑Ai},
  title        = {Auto‑Pentest‑GPT‑AI: LLM Powered Pentesting for your software},
  howpublished = {\url{https://github.com/Armur‑Ai/Auto‑Pentest‑GPT‑AI}},
  note         = {Accessed: 2025‑10‑16},
  year         = {2025}
}

@misc{XBOW,
  title        = {XBOW: AI‑Powered Penetration Testing Platform},
  howpublished = {\url{https://xbow.com/}},
  note         = {Accessed: 2025-10-16},
  year         = {2025},
  organization = {XBOW USA Inc.},
  key = {XBOW USA Inc.}
}

@inproceedings{huang2023penheal,
  title={Penheal: A two-stage llm framework for automated pentesting and optimal remediation},
  author={Huang, Junjie and Zhu, Quanyan},
  booktitle={Proceedings of the workshop on autonomous cybersecurity},
  pages={11--22},
  year={2023}
}

@article{xu2024autoattacker,
  title={Autoattacker: A large language model guided system to implement automatic cyber-attacks},
  author={Xu, Jiacen and Stokes, Jack W and McDonald, Geoff and Bai, Xuesong and Marshall, David and Wang, Siyue and Swaminathan, Adith and Li, Zhou},
  journal={arXiv preprint arXiv:2403.01038},
  year={2024}
}

@article{ginige2025autopentester,
  title={AutoPentester: An LLM Agent-based Framework for Automated Pentesting},
  author={Ginige, Yasod and Niroshan, Akila and Jain, Sajal and Seneviratne, Suranga},
  journal={arXiv preprint arXiv:2510.05605},
  year={2025}
}

@inproceedings{shen2025pentestagent,
  title={Pentestagent: Incorporating llm agents to automated penetration testing},
  author={Shen, Xiangmin and Wang, Lingzhi and Li, Zhenyuan and Chen, Yan and Zhao, Wencheng and Sun, Dawei and Wang, Jiashui and Ruan, Wei},
  booktitle={Proceedings of the 20th ACM Asia Conference on Computer and Communications Security},
  pages={375--391},
  year={2025}
}

@article{ji2024testing,
  title={Testing and understanding erroneous planning in llm agents through synthesized user inputs},
  author={Ji, Zhenlan and Wu, Daoyuan and Ma, Pingchuan and Li, Zongjie and Wang, Shuai},
  journal={arXiv preprint arXiv:2404.17833},
  year={2024}
}

@article{cao2025large,
  title={Large language models for planning: A comprehensive and systematic survey},
  author={Cao, Pengfei and Men, Tianyi and Liu, Wencan and Zhang, Jingwen and Li, Xuzhao and Lin, Xixun and Sui, Dianbo and Cao, Yanan and Liu, Kang and Zhao, Jun},
  journal={arXiv preprint arXiv:2505.19683},
  year={2025}
}

@article{blum1997fast,
  title={Fast planning through planning graph analysis},
  author={Blum, Avrim L and Furst, Merrick L},
  journal={Artificial intelligence},
  volume={90},
  number={1-2},
  pages={281--300},
  year={1997},
  publisher={Elsevier}
}

@article{mirzadeh2024gsm,
  title={Gsm-symbolic: Understanding the limitations of mathematical reasoning in large language models},
  author={Mirzadeh, Iman and Alizadeh, Keivan and Shahrokhi, Hooman and Tuzel, Oncel and Bengio, Samy and Farajtabar, Mehrdad},
  journal={arXiv preprint arXiv:2410.05229},
  year={2024}
}

@article{lin2025zebralogic,
  title={Zebralogic: On the scaling limits of llms for logical reasoning},
  author={Lin, Bill Yuchen and Bras, Ronan Le and Richardson, Kyle and Sabharwal, Ashish and Poovendran, Radha and Clark, Peter and Choi, Yejin},
  journal={arXiv preprint arXiv:2502.01100},
  year={2025}
}

@article{yamin2024failure,
  title={Failure modes of llms for causal reasoning on narratives},
  author={Yamin, Khurram and Gupta, Shantanu and Ghosal, Gaurav R and Lipton, Zachary C and Wilder, Bryan},
  journal={arXiv preprint arXiv:2410.23884},
  year={2024}
}

@article{chi2024unveiling,
  title={Unveiling causal reasoning in large language models: Reality or mirage?},
  author={Chi, Haoang and Li, He and Yang, Wenjing and Liu, Feng and Lan, Long and Ren, Xiaoguang and Liu, Tongliang and Han, Bo},
  journal={Advances in Neural Information Processing Systems},
  volume={37},
  pages={96640--96670},
  year={2024}
}

@misc{MnM_PTaaS_2024,
  author       = {MarketsandMarkets},
  title        = {Penetration Testing as a Service Market Size \& Share Analysis – Global Forecast to 2029},
  year         = {2024},
  howpublished = {\url{https://www.marketsandmarkets.com/Market-Reports/penetration-testing-as-a-service-market-36245315.html}},
  note         = {Accessed: 2025-10-03},
}

@misc{MarketsandMarkets2024,
  author       = {MarketsandMarkets},
  title        = {Penetration Testing Market Size, Size, Growth \& Latest Trends},
  year         = {2024},
  howpublished = {\url{https://www.marketsandmarkets.com/Market-Reports/penetration-testing-market-13422019.html}},
  note         = {Accessed: 2025-10-03},
}

@misc{CISA2023,
  author       = {{Cybersecurity and Infrastructure Security Agency}},
  title        = {Penetration Testing Services},
  year         = {2023},
  howpublished = {\url{https://www.cisa.gov/resources-tools/services/penetration-testing}},
  note         = {U.S. Department of Homeland Security},
}

@misc{mayoralvilches2025caiopenbugbountyready,
      title={CAI: An Open, Bug Bounty-Ready Cybersecurity AI},
      author={Víctor Mayoral-Vilches and Luis Javier Navarrete-Lozano and María Sanz-Gómez and Lidia Salas Espejo and Martiño Crespo-Álvarez and Francisco Oca-Gonzalez and Francesco Balassone and Alfonso Glera-Picón and Unai Ayucar-Carbajo and Jon Ander Ruiz-Alcalde and Stefan Rass and Martin Pinzger and Endika Gil-Uriarte},
      year={2025},
      eprint={2504.06017},
      archivePrefix={arXiv},
      primaryClass={cs.CR},
      url={https://arxiv.org/abs/2504.06017},
}

@inproceedings{deng2024pentestgpt,
  title={$\{$PentestGPT$\}$: Evaluating and harnessing large language models for automated penetration testing},
  author={Deng, Gelei and Liu, Yi and Mayoral-Vilches, V{\'\i}ctor and Liu, Peng and Li, Yuekang and Xu, Yuan and Zhang, Tianwei and Liu, Yang and Pinzger, Martin and Rass, Stefan},
  booktitle={33rd USENIX Security Symposium (USENIX Security 24)},
  pages={847--864},
  year={2024}
}

@inproceedings {depasquale_chainreactor,
author = {Giulio De Pasquale and Ilya Grishchenko and Riccardo Iesari and Gabriel Pizarro and Lorenzo Cavallaro and Christopher Kruegel and Giovanni Vigna},
title = {{ChainReactor}: Automated Privilege Escalation Chain Discovery via {AI} Planning},
booktitle = {33rd USENIX Security Symposium (USENIX Security 24)},
year = {2024},
isbn = {978-1-939133-44-1},
address = {Philadelphia, PA},
pages = {5913--5929},
url = {https://www.usenix.org/conference/usenixsecurity24/presentation/de-pasquale},
publisher = {USENIX Association},
month = aug
}

\end{document}